\documentclass[prl, showpacs, twocolumn, floatfix, superscriptaddress]{revtex4}

\usepackage{graphicx}
\usepackage{amsmath, amsfonts, amssymb, bm}

\begin{document}
\title{Strong-field spatial interference in a tailored electromagnetic bath}
\author{Mihai \surname{Macovei}}
\affiliation{Max-Planck-Institut f\"ur Kernphysik, Saupfercheckweg 1, D-69117 
Heidelberg, Germany}

\author{J\"{o}rg \surname{Evers}}
\affiliation{Max-Planck-Institut f\"ur Kernphysik, Saupfercheckweg 1, D-69117 
Heidelberg, Germany}

\author{Gao-xiang \surname{Li}}
\affiliation{Department of Physics, Huazhong Normal University, Wuhan 430079, China}
\affiliation{Max-Planck-Institut f\"ur Kernphysik, Saupfercheckweg 1, D-69117 
Heidelberg, Germany}

\author{Christoph H. \surname{Keitel}}
\affiliation{Max-Planck-Institut f\"ur Kernphysik, Saupfercheckweg 1, D-69117 
Heidelberg, Germany}
\email{keitel@mpi-hd.mpg.de}

\date{\today}
\begin{abstract}
Light scattered by a regular structure of atoms 
can exhibit interference signatures, similar to the 
classical double-slit. These first-order interferences, 
however, vanish for strong light intensities, restricting 
potential applications. Here, we show how to overcome these
limitations to quantum interference in strong fields.
First, we recover the first-order 
interference in strong fields via a tailored 
electromagnetic bath with a suitable frequency 
dependence.
At strong driving,  the optical properties
for different spectral bands are distinct, 
thus extending the set of observables. 
We further show that for a two-photon detector
as, e.g., in lithography,
increasing the field intensity leads to twice the
spatial resolution of the second-order interference pattern 
compared to the weak-field case.
%
\end{abstract}
\pacs{42.50.Hz, 42.50.St, 42.50.Ct}

%
%
%
%
%

\maketitle
If light is scattered by a structure such that different
indistinguishable pathways connect source and detector,
then interference effects may arise~\cite{BornWolf}. The classical example
is the double slit experiment, demonstrated experimentally in numerous different 
setups~\cite{otherschemes}.
A modern archetype realization employs two nearby atoms scattering
near-resonant laser light~\cite{eichmann,mirror,int1,int2,int3}.
As compared to single atom systems, the geometry of the two-particle
setup gives rise to interference phenomena in the scattered 
light. 
In particular, the beautiful experiment by Eichmann et 
al.~\cite{eichmann} has led to a discussion on the interpretation of the
first-order interference effects in terms of a double slit~\cite{int1}.
This interference, however, is restricted to 
low incident light intensity and vanishes at strong 
driving~\cite{int1,int2,int3,cbs,vogel}. 
The reason is that in the strong field limit, the two-particle 
collective dressed states are uniformly populated, such that the 
interference fringe visibility is zero. This restricts potential 
applications, as has been repeatedly reported in different areas
of optical physics. For example, coherent backscattering from 
disordered structures of atoms predominantly relies on the 
interference of coherently scattered light~\cite{cbs}, 
as well as the generation of squeezed coherent light by scattering
light off of a regular structure~\cite{vogel}.
Other applications are lithography~\cite{double-res},
where writing structures with high contrast requires a large
visibility, or precision measurements and optical information 
processing, where strong light fields may lead to increased resolution,
high signal-to-noise ratios or a rapid coherent system evolution.
More generally, the strong-field limit is desirable, as then the
different lines of the spectrum are well-separated, just as
in the Mollow resonance fluorescence spectrum of a single strongly-driven
two-level atom. The spectral separation allows for a clearer interpretation 
and gives rise to 
an extended set of observables.

Thus in this Letter, we discuss quantum interference in strong
driving fields. First, we demonstrate how to recover 
the first-order interference fringes in the strong-field case.
This is achieved by mediating the atom-laser interactions 
through a surrounding bath with  different photon mode densities 
at the various dressed-state frequencies.
Techniques to modify the vacuum as required here have been demonstrated 
in various contexts~\cite{Cexp,John}. 
The spontaneous decay rates are proportional to the density of modes at 
the transition frequencies, and thus the dressed-state populations redistribute.
As in the strong-field limit the spectral lines are well-separated, 
we are led to define optical properties for each of the spectral lines
separately, yielding an extended set of observables.
We show that by a suitable redistribution of the dressed-state populations,
full interference fringe visibility can be achieved in the 
central band of the emitted light, thus demonstrating
the recovery of the interference. Interestingly, in this case,
the scattered light is entirely coherent despite the strong driving.
%
\begin{figure}[b]
\begin{center}
\includegraphics[width=6.5cm]{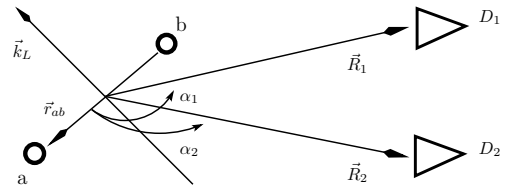}
\end{center}
\caption{\label{fig-1} Two nearby two-state emitters $a$ and $b$,
separated by $\vec r_{ab}$, and driven by a resonant strong external 
laser field with wave vector $\vec k_{L}$. Detectors $\{D_{1}, D_{2}\}$ 
are used to measure correlations among the emitted photons. 
$\{\alpha_{1}, \alpha_{2}\}$ are the angles between $\vec r_{ab}$ 
and the observation directions $\{\vec R_{1}, \vec R_{2}\}$.}
\end{figure}
%
Second, we investigate the spatial 
dependence of the second-order correlation function and focus
on the case of a single two-photon detector, as, for example, 
in lithography with a medium sensitive to two-photon exposure.  
We show that in this setup, by increasing the driving field
intensity, the spatial resolution of the
central strong-field second-order interference pattern 
can be increased by a factor of two as compared to the 
corresponding weak-field pattern.
Thus structures with high spatial resolution and signal intensity
can be created using strong driving fields.
Our scheme can be realized in a wide range of systems, and can
also be reversed to analyze the structure of the scatterers,
as discussed later.
Finally, we 
generalize our results to the case of a linear chain of $N$ atoms.

{\it The model. - } 
We first investigate a pair of distinguishable non-overlapping 
two-state emitters, $a$ and $b$, both with atomic transition 
frequency $\omega_{0}$, at positions $\vec{r}_a$, $\vec{r}_b$,
and separated by $\vec r_{ab}$.
The external laser field has frequency $\omega_{L}= ck_{L}$, 
wave vector $\vec k_{L}$, and is aligned such that 
$\vec k_{L}\cdot \vec r_{ab}=0$ (see Fig.~\ref{fig-1}).  
Our aim is to induce interference phenomena in the light scattered 
by these radiators in the intense driving field limit. We treat 
the problem in a general form, comprising any form of the environmental 
electromagnetic field (EMF) satisfying the Born-Markov conditions. 
The results will be generalized to a linear structure of
$N>2$ atoms in the final part.
The laser-dressed atomic system is described, in the electric dipole and
rotating wave approximations, by the Hamiltonian $H$ = $H_{0}$ + $H_{I}$, 
where
\begin{align}
H_{0} &= \sum_{k}\hbar(\omega_{k}-\omega_{L})a^{\dagger}_{k}a_{k} 
+ \sum_{j\in \{a,b\}}\hbar \tilde \Omega_{j} R_{zj}, \label{H0}  
\allowdisplaybreaks[2]\\
H_{I} &= i\sum_{k}\sum_{j \in \{a,b\}}(\vec g_{k}\cdot \vec d_{j})
\left \{ 
a^{\dagger}_{k} \left (R_{zj}\frac{\sin{2\theta_{j}}}{2}  
- R^{(j)}_{21}\sin^{2}{\theta_{j}}
\right . \right . 
\nonumber \\
&+ \left . \left .  R^{(j)}_{12}\cos^{2}{\theta_{j}} \right )
e^{-i(\vec k - \vec k_{L})\cdot \vec r_{j}} - \textrm{H.c.} \right \}. 
\label{HI} 
\end{align}
$H_{0}$ represents the Hamiltonian of the free EMF and free dressed 
atomic subsystems, respectively, while $H_{I}$ accounts for the 
interaction of the laser-dressed atoms with the EMF.  
$a_{k}$ and $a^{\dagger}_{k}$ are the radiation field annihilation 
and creation operators obeying the commutation relations
$[a_{k},a^{\dagger}_{k^{'}}]=\delta_{kk^{'}}$, 
and $[a_{k},a_{k^{'}}]$= $[a^{\dagger}_{k},a^{\dagger}_{k^{'}}]=0$. 
The atomic operators $R^{(j)}_{\alpha \beta}=|\tilde 
\alpha\rangle_{j} {}_{j}\langle \tilde \beta|$ describe the transitions 
between the dressed states $|\tilde \beta \rangle_{j}$ and $|\tilde 
\alpha \rangle_{j}$ in atom $j$ for $\alpha \not= \beta$ and 
dressed-state populations for $\alpha=\beta$, and satisfy the 
commutation relation $[R^{(j)}_{\alpha\beta},
R^{(l)}_{\alpha^{'}\beta^{'}}]=\delta_{jl}
[\delta_{\beta \alpha^{'}}R^{(j)}_{\alpha \beta^{'}}-\delta_{\beta^{'} 
\alpha}R^{(j)}_{\alpha^{'}\beta}]$. 
The operators $R^{(j)}_{\alpha \beta}$ 
can be represented through the bare state operators via the 
transformations $|1\rangle_{j} = \sin{\theta}|\tilde
2\rangle_{j}+\cos{\theta}|\tilde 1\rangle_{j}$ and
$|2\rangle_{j}=\cos{\theta}|\tilde 2\rangle_{j}-\sin{\theta}
|\tilde 1\rangle_{j}$. 
We define $R_{zj}=|\tilde 2\rangle_j {}_j\langle \tilde 2|
- |\tilde 1\rangle_j {}_j\langle \tilde 1|$.
Further, $\tilde \Omega=\tilde \Omega_j=
[\Omega^{2}+(\Delta/2)^{2}]^{1/2}$ is the 
generalized Rabi frequency, with 
$2\Omega=(\vec d\cdot \vec E_{L})/\hbar$. Here, $E_{L}$ is the electric 
laser field strength, and $\vec d \equiv \vec d_{a} = \vec d_{b}$ is the 
transition dipole matrix element. 
The detuning $\Delta = \omega_{0} - \omega_{L}$ is characterized by
$\cot{2\theta} = \Delta/(2\Omega)$.
The dressed state transition frequencies are $\omega_L, 
\omega_\pm = \omega_L \pm 2\tilde{\Omega}$.

The two-particle spontaneous decay and the vacuum-mediated 
collective interactions are given by the frequency-dependent 
expression 
\begin{eqnarray}
\gamma_{jl}(\omega) &=& \hbar^{-2}\sum_{k} (\vec g_{k}\cdot \vec d)^2
e^{i\vec k \cdot \vec r_{jl}} 
\Theta(\omega_{k},\omega) 
\nonumber \\
 &=& \gamma(\omega)[\chi_{jl}(\omega)+i\Omega_{jl}(\omega)]\,.
\end{eqnarray}
The coupling constant between atom and environment
is $g_{k}$, while $\Theta(\omega_{k},\omega)$ defines  the structure of
the Markovian bath. Independent 
of the atom-vacuum coupling, the collective parameters $\chi_{jl}$ and 
$\Omega_{jl}$ ($j\neq l$) tend to zero in the large-distance case 
$r_{jl} \to \infty$ which  corresponds to the absence of coupling 
among the emitters. 

{\it Intensities. - } 
Driving a single two-state atom with a strong near-resonant
laser field splits the 
resonance fluorescence spectrum into the well-known
three-peaked Mollow spectrum.
Then it is reasonable to consider optical properties
for each spectral band separately. 
A similar splitting occurs in a two-atom system.
If the laser beam is 
perpendicular to the line connecting the atoms, i.e. 
$\vec k_{L}\cdot \vec r_{ab}=0$, then the central ($CB$) 
and left/right sideband ($LB/RB$) spectral
intensities are given by
\begin{subequations}
\label{Ints}
\begin{align}
I_{CB}(\vec R_1)=&\frac{1}{4}\sum_{j,l \in \{a,b\}}\Psi_{R_1}
(\vec r_{jl},\omega_{L})\langle R_{zj}R_{zl}\rangle \sin^{2}{2\theta}\,, 
\allowdisplaybreaks[2]  \\
I_{LB}(\vec R_1)=&\sum_{j,l \in \{a,b\}}\Psi_{R_1}(\vec r_{jl},\omega_{-})
\langle R^{(j)}_{12}R^{(l)}_{21}\rangle \sin^{4}{\theta}\,, 
\allowdisplaybreaks[2]  \\
I_{RB}(\vec R_1)=&\sum_{j,l \in \{a,b\}}\Psi_{R_1}(\vec r_{jl}, \omega_{+})
\langle R^{(j)}_{21}R^{(l)}_{12}\rangle \cos^{4}{\theta}\,. 
\end{align}
\end{subequations}
Here, $\Psi_{R_1}(\vec r_{jl},\omega)=\Psi_{R_1}(\omega)\exp{[i(\omega/c)r_{jl}
\cos{\alpha_1}]}$ with angle $\alpha_1$  between observation 
direction $\vec R_1$ and distance vector $\vec r_{jl}$. 
This pre-factor  depends on the atom-environment coupling and 
in general the frequency with $R_1=|\vec R_1| \gg k^{-1}_{L}$. 
We have assumed the strong field limit 
$\tilde \Omega \gg \{\gamma(\omega_{+}), \gamma(\omega_{-}),
 \gamma(\omega_{L})\}$ 
and $\tilde \Omega \gg \{\gamma(\omega_{+})\Omega_{ab}(\omega_{+}), 
\gamma(\omega_{-})\Omega_{ab}(\omega_{-}), \gamma(\omega_{L})
\Omega_{ab}(\omega_{L})\}$ in deriving Eqs.~(\ref{Ints}).
Note that multi-atom systems may exhibit additional spectral 
line splittings due to the dipole-dipole interaction between the 
emitters, if the interatomic distance is small compared to the
transition wavelength.
%

{\it Visibilities. - } If the inter-particle separation is large enough
to ignore the line-splittings caused by dipole-dipole interactions,
then the visibilities $V=[I_{max} - I_{min}]/[I_{max} + I_{min}]$ 
for each of the central, left and right spectral band, 
follow from Eqs.~(\ref{Ints}) as
\begin{subequations}
\label{Vi}
\begin{eqnarray}
V_{CB} &=& \sigma_{ee} + \sigma_{gg} - \sigma_{ss} -\sigma_{aa} \,, 
 \\ 
V_{LB(RB)} &=& [\sigma_{ss} - \sigma_{aa}]/[1 \mp \sigma_{ee} \pm 
\sigma_{gg}]\,. 
\end{eqnarray}
\end{subequations}
We have introduced in Eq.~(\ref{Vi}) the two-atom collective dressed 
states as $|\Psi_{e}\rangle=|\tilde 2_{a},\tilde 2_{b}\rangle$, 
$|\Psi_{s(a)}\rangle=\{|\tilde 2_{a},\tilde 1_{b}\rangle \pm |\tilde 2_{b},
\tilde 1_{a}\rangle \}/\sqrt{2}$, $|\Psi_{g}\rangle=|\tilde 1_{a},
\tilde 1_{b}\rangle$. The expectation values 
$\sigma_{\alpha \beta}=\langle |\Psi_{\alpha}
\rangle \langle \Psi_{\beta}|\rangle$ describe the corresponding transitions 
($\alpha \not = \beta$) and populations ($\alpha = \beta$)
$(\{\alpha,\beta \} \in \{e,s,a,g \})$. 
If the population is now transferred into a particular collective 
dressed state, then the spectral band visibilities will behave according to 
Eqs.~(\ref{Vi}). One can easily observe that all visibilities vanish if
the atomic population is uniformly distributed over two-particle 
collective dressed states, recovering previous results~\cite{int1,int2,int3}.

{\it Two-particle quantum dynamics. - } 
Introducing the notations 
$x=2(\sigma_{ee}-\sigma_{gg})$, 
$y=\sigma_{ss}-\sigma_{aa}$, and 
$z=\sigma_{ee}+\sigma_{gg}-\sigma_{ss}-\sigma_{aa}\equiv V_{CB}$,
the dressed-state atomic correlators in 
Eqs.~(\ref{Ints}, \ref{Vi}) follow from Eqs.~(\ref{H0},\ref{HI}) as~\cite{mario}
\begin{subequations}
\label{twoEq}
\begin{align}
\dot x(t) &= -2\xi^{(+)}x + 4\zeta^{(-)}_{ab}y + 4 \xi^{(-)} \,, 
 \allowdisplaybreaks[2]\\
\dot y(t) &= -\zeta^{(-)}_{ab}x - 2 (c^{(0)}_{ab} + \xi^{(+)})
y + 2\zeta^{(+)}_{ab}z \,, 
 \allowdisplaybreaks[2] \\
\dot z(t) &= 2\xi^{(-)}x + 4\zeta^{(+)}_{ab}y - 4\xi^{(+)}z \,. 
\end{align}
\end{subequations}
%
The coefficients are 
$\xi^{(\pm)}$ = $\gamma(\omega_{-})
\sin^{4}{\theta} \pm \gamma(\omega_{+})\cos^{4}{\theta}$,
$\zeta^{(\pm)}_{ab} = \gamma(\omega_{-})\chi_{ab}(\omega_{-})
\sin^{4}{\theta} \pm \gamma(\omega_{+})\chi_{ab}(\omega_{+})\cos^{4}
{\theta}$, 
and 
$c^{(0)}_{ab} = \gamma(\omega_{L})[1 - \chi_{ab}
(\omega_{L})]\sin^{2}{2\theta}$. 
Simple analytical expressions for the two-atom steady-state quantum 
dynamics can be obtained in some particular cases. For example, 
for the large-distance case $k_L r_{ab}\gg 2\pi$ one obtains
$\zeta^{(\pm)}_{ab}\to 0$ and $c^{(0)}_{ab} \to \gamma(\omega_{L})\sin^{2}{2\theta}$,
such that
%
\begin{eqnarray}
x = 2\xi^{(-)}/\xi^{(+)}, \quad y = 0, \quad {\rm and }~~z = [\xi^{(-)}/
\xi^{(+)}]^{2}. \label{small}
\end{eqnarray}
In this case, the diagonal atomic dynamics is independent 
of the inter-atomic separation. 

{\it First-order interference pattern. -} For resonant 
driving ($\theta = \pi/4$), 
the atomic population is equally distributed over the two-atom 
dressed states resulting in the absence of an interference 
pattern. Only small values for $V_{CB}$ can be obtained for 
off-resonant pumping, while $V_{LB}$ and $V_{RB}$ are always 
zero [see Eqs.~(\ref{Vi}) and (\ref{small})]. A weak interference 
pattern can be induced in the sidebands  if the splitting of the 
dressed levels is large such that $2\tilde \Omega/
\omega_{0} \ll 1$ is not negligible. Then the dressed
states couple differently to the EMF, i.e.,
$\gamma(\omega_{-})\chi_{ab}(\omega_{-}) \not = \gamma(\omega_{+})
\chi_{ab}(\omega_{+}) \not = 0$~\cite{mario}. Nevertheless, 
for practical situations, the first-order  interference  
vanishes for a strongly driven atomic pair in free space.
\begin{figure}[t]
\begin{center}
\includegraphics[width=7.5cm]{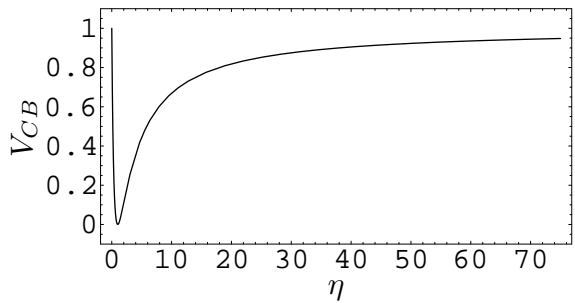}
\end{center}
\caption{\label{fig-2} Central band visibility $V_{CB}$ as function of 
$\eta$ for $\theta = \pi/4$ and $k_{L}r_{ab} \gg 2\pi$.}
\end{figure}

In the following, we show how the interference can be recovered
for two strongly driven atoms by modifying the surrounding electromagnetic 
reservoir.  In effect, this alters the parameter 
$\eta = \gamma(\omega_{+})/\gamma(\omega_{-})$, which in free space
is 1.
We assume the driving fields to be on resonance, 
i.e., $\theta = \pi/4$, with $\Omega \gg \{\gamma(\omega_\pm), \gamma(\omega_L)\}$. 
Figure~\ref{fig-2} shows the dependence of the central spectral band 
visibility, 
$V_{CB}=[(1-\eta)/(1+\eta)]^{2}$, versus the ratio 
$\eta$. 
%
\begin{figure}[t]
\begin{center}
\includegraphics[width=7.5cm]{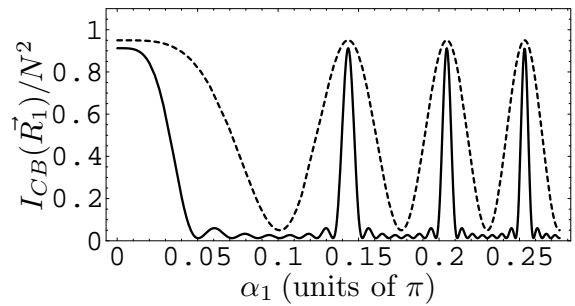}
\end{center}
\caption{\label{fig-3} Central band intensity $I_{CB}(\vec R_1)/N^{2}$ 
[in units of $\Psi_{R_1}(\omega_{L})/4$] as function of $\alpha_1$. 
Here $\theta = \pi/4$, $k_{L}r_{ab}=20\pi$ and $V_{CB}=0.9$. Solid 
line: $N=8$, dashed curve: $N=2$.}
\end{figure}
%
Maximum visibility ($V_{CB}=1$) can be obtained for $\eta \ll 1$ or $\eta 
\gg 1$. This corresponds to an interference pattern with a bright center.
In these cases, $\sigma_{ee} \to 1$ ($\eta \ll 1$) 
or $\sigma_{gg} \to 1$ ($\eta \gg 1$), while the other two collective dressed 
states are empty ($\sigma_{ss}=\sigma_{aa}=0$). Thus, $V_{LB}=V_{RB}=0$. 
In other words, if the densities of the EMF modes at 
the dressed transition frequencies 
$\omega_{\pm} = \omega_{L} \pm 2\tilde \Omega$ 
differ considerably, then $\eta \ll 1$ or $\eta \gg 1$, and the 
interference pattern is recovered in the central band.
Figure~\ref{fig-3} shows a corresponding interference 
pattern versus detection angle $\alpha$.
We now analyze the interference in terms of scattering via 
symmetric and anti-symmetric collective
states~\cite{int2}. Transitions involving symmetric [anti-symmetric]
collective states give rise to interference with a bright [dark] center.
If both channels have equal probability, the interference fringes
wash out. In our system for $\eta\ll 1$ or $\eta\gg 1$, however, only 
symmetric collective states are populated, as can be seen from 
the definition of $|\Psi_e\rangle$ and $|\Psi_g\rangle$. Thus we always find
bright center interference. Note that independent of the 
collective state, the single atom dressed states can be either
symmetric ($|\tilde 2\rangle$) or anti-symmetric ($|\tilde 1\rangle$).

Our scheme can, for example, be implemented by trapping the 
scatterers in a cavity with suitable frequency dependence. 
Experiments on modifying the single-atom 
Mollow spectrum in cavities have already been reported~\cite{Cexp}. Other 
methods are to embed the two particles in a photonic band-gap
material~\cite{John}, or to additionally pump the dressed-atom sample with 
chaotic fields~\cite{chaot}. 
The experimental control of  population in an artificial 
two-state atom~\cite{Eart} suggests a possible realization in mesoscopic
systems. Our results can also be used to analyze the structure
of the scattering material, e.g., the geometric distribution of the
scatterers, in particular using the extension to many scatterers discussed below.
%

{\it Second-order correlation functions. - } We now turn to the 
second-order correlation function of the resonance fluorescence
emitted in the three spectral bands, 
\begin{subequations}
\label{g2}
\begin{align}
g^{(2)}_{CB}(\vec R_{1},\vec R_{2}) = 1 + \frac{(1-V^{2}_{CB})
\cos{\delta^{(0)}_{1}}\cos{\delta^{(0)}_{2}}}
{D_{CB}} \,, 
\allowdisplaybreaks[2] \\
g^{(2)}_{LB,RB}(\vec R_{1},\vec R_{2}) = \frac{p_{L,R}\: [1 + 
\cos(\delta^{(\mp)}_{1}-\delta^{(\mp)}_{2})] }{D_{LB,RB}} \,.
\end{align}
\end{subequations}
Here, $D_n = \prod_{m \in \{1,2 \}}\:
[1+V_{n}\cos{\delta^{(n)}_{m}}]$, and
$p_{L}=2\sigma_{gg}/[1-\sigma_{ee}+\sigma_{gg}]^{2}$ and 
$p_{R}=2\sigma_{ee}/[1+\sigma_{ee}-\sigma_{gg}]^{2}$. The
argument $\delta_{m}^{(n)}=k_n r_{ab}\cos{\alpha_{m}}$ 
($m\in\{1,2\}$) is evaluated at frequency
$\{\omega_{L}, \omega_{-}, \omega_{+}\}$ with wavevector $k_n$ for 
the respective $n$-values $\{0 = CB, -= LB, + = RB\}$. 

For $V_{CB}=1$, one finds $g^{(2)}_{CB}(\vec R_{1},\vec R_{2})=1$,
i.e., fully coherent light.
%
\begin{figure}[t]
\begin{center}
\includegraphics[width=8cm]{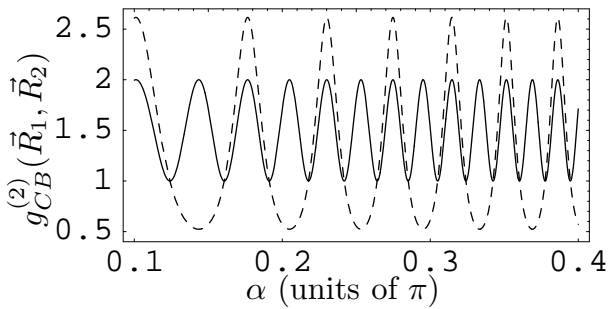}
\end{center}
\caption{\label{fig-4} Central band second-order correlation function  
$g_{CB}^{(2)}(\vec{R}_1,\vec{R}_2 )$. The solid line depicts the 
strong-field limit ($V_{CB}=0$) while the dashed curve describes 
the weak-field case with $\Omega/\gamma =0.9$. Here $\Delta=0$, 
$k_{L}r_{ab}=20\pi$ and $\vec R_{1}=\vec R_{2}$.}
\end{figure}
%
In the absence of first-order interference ($V_{CB,LB,RB}=0)$, 
in the standard vacuum, the second-order 
correlation functions do exhibit interference:
\begin{subequations}
\label{noVi}
\begin{eqnarray}
g^{(2)}_{CB}(\vec R_{1},\vec R_{2}) &=& 1 + \cos{\delta_{1}}\cos{\delta_{2}} 
\,,  \\
g^{(2)}_{LB,RB}(\vec R_{1},\vec R_{2}) &=& [1 + \cos(\delta_{1}-\delta_{2})]/2 \,.
\end{eqnarray}
\end{subequations}

Our main interest here is the second-order spatial interference 
resolution~\cite{double-res} for a two-photon detector with
$\delta_{1}=\delta_{2}\equiv \delta$, e.g., a medium
sensitive to two-photon exposure.
In the strong-field limit $\Omega/\gamma \gg 1$, from 
Eq.~(\ref{noVi}) we 
find $g^{(2)}_{CB}(\vec R) = 1+\cos^{2}{\delta}$. 
In the weak field case $\Omega/\gamma <1$ without spectral band 
separation, however, one finds $g^{(2)}(\vec R) = [s/(s + \cos{\delta})]^{2}$
with $s=1 + 2(\Omega/\gamma)^{2}$~\cite{int2}. 
Remarkably, increasing the driving
field strength effectively doubles the spatial fringe resolution in the
central spectral band in this detector setup relevant, e.g., to 
lithography. This is illustrated in Fig.~\ref{fig-4},
which shows both cases as function of the detector positions. 
Thus high-resolution spatial structures can be achieved
using strong driving fields.
We note in passing that sub-Poissonian or Poissonian photon statistics~\cite{walls} 
can be generated in all three spectral lines,
and super-Poissonian photon statistics in the central band, see, e.g., 
Eq.~(\ref{noVi}).

{\it Multi-atom sample. - } We now extend our first-order interference
analysis to a 
many-atom ensemble. If $N$ independent two-level emitters are uniformly 
distributed in a linear chain ($r_{ab}=r_{i,i-1}$), then their central-band 
intensity [in units of $\sin^{2}({2\theta})\Psi_{R}(\omega_{L})/4$] 
evaluates to
\begin{align}
I_{CB}(\vec R_1) =& N(1-z) + z \mathcal{F}(\delta_1) \,.
\end{align}
Here, $\delta_i = k_{L}r_{ab}\cos{\alpha_i}$,
$\mathcal{F}(x) = \sin^2[Nx/2] / \sin^2[x/2]$,
and $z$ is given in Eq.~(\ref{small}). Maxima of $I_{CB}$ 
occur for $k_{L}r_{ab}\cos{\alpha}=2\pi n$  with $z=1$,
where $I^{(max)}_{CB}(\vec R) \propto N^{2}$. 
Thus the central-band visibility is significantly improved 
(see Fig.~{\ref{fig-3}}), while the sub-wavelength pattern resolution 
scales with the atom number~\cite{BornWolf}. 

In summary, it was shown how first-order interference can be recovered
in the fluorescence light of strongly driven atoms.
The key idea is to modify the surrounding electromagnetic 
vacuum, giving rise to a redistribution of the collective dressed 
state populations. Under strong driving, visibilities have to be defined
for each emitted spectral band separately, providing an extended
set of observables. Finally, the second-order interference fringes 
for two-photon detection have double resolution in the strong-field case 
as compared to the weak-field pattern.


\end{document}